\documentclass[showpacs,preprintnumbers,amsmath,amssymb,nofootinbib]{revtex4}
\usepackage{graphicx}
\usepackage{dcolumn}
\usepackage{bm}
\usepackage{multirow}
\usepackage{subfigure}

\begin{document}
\title{Constraint on the light quark mass $m_q$ from QCD Sum Rules in the $I=0$ scalar channel}

\author{Jia-Min Yuan$^1$}
\author{Zhu-Feng Zhang$^{1,2}$}
\author{T. G. Steele$^2$}
\author{Hong-Ying Jin$^3$}
\author{Zhuo-Ran Huang$^3$}

\affiliation{
$^1$Physics Department, Ningbo University, Zhejiang Province, 315211, P. R. China
\\
$^2$Department of Physics and Engineering Physics, University of Saskatchewan, Saskatoon,
Saskatchewan, S7N 5E2, Canada
\\
$^3$Zhejiang Institute of Modern Physics, Zhejiang University, Zhejiang Province, 310027, P. R. China
}

\begin{abstract}
In this paper, we reanalyze the $I=0$ scalar channel with the improved Monte-Carlo based QCD sum rules,
which combines the rigorous H\"older-inequality-determined sum rule window and a two Breit-Wigner type
resonances parametrization for the phenomenological spectral density
that satisfies the the low-energy theorem for the scalar form factor.
Considering the uncertainties of the QCD parameters and the experimental masses and widths of
the scalar resonances $\sigma$ and $f_0(980)$, we obtain a prediction for light quark mass
$m_q(2\,\textrm{GeV})=\frac{1}{2}(m_u(2\,\textrm{GeV})+m_d(2\,\textrm{GeV}))=4.7^{+0.8}_{-0.7}\,\textrm{MeV}$, 
which is consistent with the PDG (Particle Data Group) value and QCD sum rule determinations in the pseudoscalar channel.
This agreement provides a consistent framework connecting QCD sum rules and low-energy hadronic physics. 
We also obtain the decay constants of $\sigma$ and $f_0(980)$ at 2\,GeV, which are
approximately $0.64-0.83$\,GeV and $0.40-0.48$\,GeV respectively.
\end{abstract}


\maketitle

\section{Introduction}

The light quark masses are fundamental parameters in QCD, thus it is important to determine these
parameters from different methods. Due to the color confinement, the light quark masses can not be
measured from experiments directly. Therefore, their values are determined by relating
the light quark masses to other physical quantities which can be obtained from theories or experiments.
The main QCD-based methods for determining the light quark masses are lattice QCD (see e.g., Ref.~\cite{Aoki:2013ldr}
for a review) and QCD sum rules (QCDSR) 
\cite{Becchi:1980vz,Bijnens:1994ci,Narison:1999uj,Narison:2002hk,Narison:2005ny,Dominguez:2008tt,Cherry:2001cj}.

The pion channel is the most common method to determine the light quark masses from QCDSR.
In Ref.~\cite{Bijnens:1994ci}, Bijnens et al.\ studied the value of the light quark mass combination
$m_u+m_d$ in QCD using both Finite Energy Sum Rules (FESR) and Laplace Sum rules (LSR) for the
divergence of the axial current with the quantum numbers of the pion,  finding
 $m_u(1\,\textrm{GeV})+m_d(1\,\textrm{GeV})=12\pm2.5\,\textrm{MeV}$, which leads to a
light quark mass $m_q(2\,\textrm{GeV})=\frac{1}{2}(m_u(2\,\textrm{GeV})+m_d(2\,\textrm{GeV}))
=4.8\pm1.0\,\textrm{MeV}$ at the Particle Data Group (PDG) standard energy scale 2\,GeV.
Later, after including five-loop order and higher order quark-mass corrections to the correlation
function of the same current, a more accurate result $m_q(2\,\textrm{GeV})=4.1\pm0.2\,\textrm{MeV}$
was found by using FESR \cite{Dominguez:2008tt}.

In addition to the divergence of the axial current, one can also relate the light quark masses to
other currents. It is clearly important to establish the self-consistency of the quark mass 
extracted from different channels. In Ref.~\cite{Cherry:2001cj}, Cherry et al.\ used the 
$I=0$ scalar current to study this problem. By linking the phenomenological spectral density 
to the $\pi\pi$ scattering amplitude, they obtained the average light quark mass 
$m_q(1\,\textrm{GeV})=5.2\pm0.6\,\textrm{MeV}$. However, the main uncertainty in this analysis 
is determining the normalization between the theoretical and phenomenological spectral density.
As discussed in Ref.~\cite{Narison:2002hk}, it is difficult to assess the hadronic uncertainties 
in Ref.~\cite{Cherry:2001cj}, motivating our alternative approach.
In this paper, we will reinvestigate the $I=0$ scalar channel using the improved Monte-Carlo based
QCD sum rule methodology recently proposed in Ref.~\cite{Wang:2016sdt}. After introducing a two Breit-Wigner
type resonances parametrization for the phenomenological spectral density normalized by the low-energy theorem, 
a Monte-Carlo based analysis will be presented for the QCD sum rule master equation with the $I=0$ 
scalar current in the rigorous H\"older-inequality-determined sum rule window. Based on this analysis, we 
will give robust constraint on the light quark mass $m_q$ and predictions for the decay constants 
of $\sigma$ and $f_0(980)$.

\section{QCD sum rule for $I=0$ scalar channel}

We consider the correlation function
\begin{equation}
\label{eq:correlator}
\Pi(q^{2})=i\int d^4x \,e^{i qx}\langle 0|T{j_s(x)j_s^\dagger(0)}|0\rangle,
\end{equation}
where $j_s=m_q \frac{1}{\sqrt{2}} (\bar u u+\bar d d)$ is the $I=0$ renormalization group invariant
scalar current and $m_q=\frac{1}{2}(m_u+m_d)$ is the average mass of $u$ and $d$ quarks. 
The theoretical representation of this function has been calculated by using the operator
product expansion (OPE) method \cite{Reinders:1981ww,Reinders:1984sr,Surguladze:1990sp},
however, it is believed that other nonperturbative contributions to the correlation function
must be included, and thus we also should include instanton contribution
$\Pi^{\textrm{(inst)}}(q^2)$ in the theoretical representation of the correlation function
\cite{Shuryak:1982qx,Elias:1998bq,Elias:1998fs,Shi:1999hm,Kisslinger:2001pk}.

To obtain a QCD sum rule, we first need to Borel-transform the theoretical representation
of the correlation function, which gives
\cite{Reinders:1981ww,Reinders:1984sr,Surguladze:1990sp,Shuryak:1982qx,Elias:1998bq,Elias:1998fs,Shi:1999hm,Kisslinger:2001pk}
\begin{equation}
\label{eq:theoretical}
\begin{split}
R^{\textrm{(theo)}}(\tau,\hat m_q)=&\frac{1}{\tau}\hat B \Pi^{\textrm{(OPE)}}(q^2)
+\frac{1}{\tau}\hat B \Pi^{\textrm{(inst)}}(q^2)=m_q^2(1/\sqrt \tau)\cdot
\left\{\frac{3}{8\pi^2}
\left(1+\frac{17}{3}\frac{\alpha_s(1/\tau)}{\pi}\right)\frac{1}{\tau^2}\right.\\
&+\frac{3}{8\pi^2}\frac{\alpha_s(1/\tau)}{\pi} \frac{2}{\tau^2}(\gamma_E-1)
+\frac{\langle\alpha_s G^2\rangle}{8\pi}
\left(1+\frac{11}{2}\frac{\alpha_s(1/\tau)}{\pi}\right)
+ 3\langle m_q\bar qq\rangle \left(1+\frac{13}{3}\frac{\alpha_s(1/\tau)}{\pi}\right)\\
&-\frac{176}{27}
\pi \kappa \alpha_s\langle \bar qq\rangle^2 \left[\frac{\alpha_s(1/\tau)}{\alpha_s(\mu^2_0)}\right]^{1/9}  \tau
+\left.\frac{3}{8\pi^2}\frac{e^{\frac{-\rho^2}{2\tau}}\rho^2}{\tau^3}\left(K_0\left(\frac{\rho^2}{2 \tau}\right)
+K_1\left(\frac{\rho^2}{2\tau}\right)\right)\right\},
\end{split}
\end{equation}
where $\hat B$ is the Borel transformation operator, $\alpha_s(1/\tau)=4\pi/(9\ln(1/(\tau \Lambda^2_{\textrm{QCD}})))$
is the running coupling constant for three flavors at scale $1/\sqrt \tau$ (the QCD scale
$\Lambda_{\textrm{QCD}}=0.353\,\textrm{GeV}$ \cite{Narison:2009vy}),
$\kappa$ is the vacuum factorization violation factor which parameterizes the deviation of the four-quark
condensate from a product of two-quark condensates, $\rho$ is the instanton size in the instanton
liquid model, and $K_0$ and $K_1$ are modified Bessel functions. We have considered the renormalization-group
(RG) improvement of the sum rules \cite{Narison:1981ts} and anomalous dimensions
for condensates \cite{Shifman:1978bx,Shifman:1978by} in Eq.~\eqref{eq:theoretical}, where $\mu_0$ is
the renormalization scale for condensates, and
\begin{equation}
m_q(1/\sqrt \tau)=\hat m_q\cdot \left[\frac{4\pi}{9\ln(\frac{1}{\tau\Lambda^2_{\textrm{QCD}}})}
\left(1-\frac{64}{81}\frac{\ln(\ln(\frac{1}{\tau \Lambda^2_{\textrm{QCD}}}))}{\ln(\frac{1}{\tau \Lambda_{\textrm{QCD}}})}\right)\right]^{4/9}
\end{equation}
is the running light quark mass at scale $1/\sqrt \tau$ where $\hat m_q$ is the RG-invariant
light quark mass. In Eq.~\eqref{eq:theoretical}, we also have included the $\alpha_s$ corrections
to dimension-4 operators, which may play an important role in the determination of the QCD sum rule 
window from the H\"older inequality as in Ref.~\cite{Wang:2016sdt}.

It is also necessary to construct a phenomenological spectral density model which is related to the correlation function through the dispersion
relation integral. Considering the resonance nature of scalar mesons, we insert the lowest two-pion intermediate
state \footnote{There exist higher intermediate states which contain more particles, 
e.g., four-pion intermediate state. However, multiple particle intermediate 
states would be kinetic suppressed by small phase space factors, 
thus we will classify these intermediate states together with other two 
particle intermediate states into ``other intermediate states'' below.}, 
as part of a complete set, into Eq.~\eqref{eq:correlator}, i.e., by inserting
$\int\frac{d^3 {\bm k_1}}{(2\pi)^3 2 E_{\bm{k_1}}}
\frac{d^3{\bm k_2}}{(2\pi)^3 2 E_{\bm{k_2}}} (|\pi^+(k_1)\pi^-(k_2)\rangle \langle \pi^+(k_1)\pi^-(k_2)|
+\frac{1}{2!}|\pi^0(k_1)\pi^0(k_2)\rangle \langle \pi^0(k_1)\pi^0(k_2)|) + \textrm{``other intermediate states''}$
for the correlation function of current $j_s$, and using Cutkosky's cutting rules \cite{Cutkosky:1960sp}, then
the phenomenological expression for ${\rm Im}\Pi(s)$ can be found:
\begin{equation}
\label{eq:formfactor}
{\rm Im}\Pi^{\textrm{(phen)}}(s)=\frac{3}{64\pi}\sqrt{1-\frac{4m_\pi^2}{s}}|F_s(s)|^2
+\textrm{contributions from excited states and continuum (ESC)},
\end{equation}
where $m_\pi$ is the mass of pion, and
$\langle 0|j_s(0)|\pi^+(k_1)\pi^-(k_2)\rangle =\frac{1}{\sqrt 2} F_s((k_1+k_2)^2)$
has been used. We have classified all contributions from intermediate states other than two-pion
intermediate state, including these from four-pion intermediate state, into contributions from ESC.
According to chiral perturbative theory (ChPT), the scalar form factor $F_s(s)$ will be normalized
by a low-energy theorem $F_s(0)=m_\pi^2$ \cite{Gasser:1990bv}, so we will constrain our phenomenological
spectral density with this condition in the following.

In Ref.~\cite{Cherry:2001cj}, the phenomenological spectral density for the $I=0$ scalar channel 
is related to the $\pi\pi$ scattering amplitude via the scalar form factor $F_s(s)$. However, because of a 
lack of experimental data which are consistent with ChPT at some energy scale, Cherry et al.~introduced
multiple assumptions for their phenomenological spectral density, which dominated the uncertainties in their analysis.
In this paper, we will perform an independent analysis by parameterizing the phenomenological spectral density with the mass spectrum 
for the $I=0$ scalar channel directly and incorporate the ChPT low-energy theorem.

The $0^+(0^{++})$ meson spectrum are rather crowded, there are too many particles with quantum
numbers $0^+(0^{++})$ listed in the Review of Particle Physics \cite{Olive:2016xmw} for a single nonet.
Many different models have been used to describe the structures of these scalar mesons in QCDSR,
including ordinary $\bar qq$ meson, four-quark state, glueball and hybrid
\cite{Reinders:1981ww,Reinders:1984sr,Narison:2000dh,Brito:2004tv,Bagan:1990sy,Forkel:2000fd,Zhang:2011qza}.
However, the possible mixing between mesons with the same quantum numbers make this problem
even more complex, and a widely accepted conclusion of research on the structures of these scalar mesons has
not been achieved.

Amongst all these $I=0$ scalar mesons, we notice that both $\sigma$ and $f_0(980)$ have the two-pion decay
mode as their dominant decay mode. Thus we can conjecture that there are contributions from poles of
$\sigma$ and $f_0(980)$ in the two-pion scalar form factor, i.e.,
$F_s(s)$ may have two poles at $s=m_\sigma-i m_\sigma \Gamma_\sigma$ and $s=m_{f_0}-i m_{f_0} \Gamma_{f_0}$,
where $m_\sigma$ and $\Gamma_\sigma$ ($m_{f_0}$ and $\Gamma_{f_0}$) are the mass and width of $\sigma$
($f_0(980)$) meson respectively.

Considering the normalization of the form factor $|F_s(0)|^2=m_\pi^4$ from ChPT, we can construct a
two Breit-Wigner type resonances model for the phenomenological spectral density which meets the
above requirements as follows \footnote{Notice that our model does not exclude other $0^+(0^{++})$
mesons from having $\bar qq$-component, however, the contributions
to the two-pion scalar form factor originate from heavier scalar mesons
should be negligible because of the exponential suppression factor in the
Borel-transformed dispersion relation integral and the form factor will be 
suppressed by the small branching ratio of the two-pion decay mode.}
\begin{equation}
\begin{split}
\label{eq:peak}
&\frac{1}{\pi}{\rm Im}\Pi^{{\rm (resonance)}}(s)=\frac{3}{64\pi^2}\cdot|F_s(s)|^2\\
=&\frac{3}{64\pi^2}\cdot m_\pi^4 \left(\beta\cdot\frac{m_\sigma^4+m_\sigma^2\Gamma_\sigma^2}{(s-m_\sigma^2)^2+m_\sigma^2\Gamma_\sigma^2}
+(1-\beta)\cdot\frac{m_{f_0}^4+m_{f_0}^2\Gamma_{f_0}^2}{(s-m_{f_0}^2)^2+m_{f_0}^2\Gamma_{f_0}^2}\right),
\end{split}
\end{equation}
where we have omitted the small mass of pion ($m_\pi=0.139\,\textrm{GeV}$ \cite{Olive:2016xmw})
in the square root in Eq.~\eqref{eq:formfactor}. The parameter $\beta$ ($0\leq\beta\leq1$) describes
the relative contribution of $\sigma$ and $f_0(980)$ to the phenomenological spectral density
in our model.

For the ESC contributions in the phenomenological spectral density, we still use the traditional
ESC model, i.e.,
\begin{equation}
\label{eq:esc}
\frac{1}{\pi}{\rm Im}\Pi^{\textrm{(ESC)}}(s)=m_q^2(1/\sqrt \tau)\cdot
\left(\frac{3}{8\pi^2}\left(1+\frac{17}{3}\frac{\alpha_s}{\pi}\right)s-\frac{3}{4\pi^2}\frac{\alpha_s}{\pi}s\ln(s\tau)
-\frac{3}{4\pi} s J_1(\sqrt{s}\rho) Y_1(\sqrt{s}\rho)\right)\theta(s-s_0),
\end{equation}
where $s_0$ is the continuum threshold separating the contributions from excited states and continuum,
$J_1$ and $Y_1$ are Bessel function of the first and second kind respectively.

Collecting Eq.~\eqref{eq:peak} and \eqref{eq:esc} together, we can obtain our phenomenological
spectral density as follows
\begin{equation}
\label{eq:model1}
\frac{1}{\pi}{\rm Im}\Pi^{\textrm{(phen)}}(s)=\frac{1}{\pi}{\rm Im}\Pi^{\textrm{(resonance)}}(s)
+\frac{1}{\pi}{\rm Im}\Pi^{\textrm{(ESC)}}(s).
\end{equation}
Then the phenomenological representation for the Borel-transformed correlation function can be obtained by
using the dispersion relation:
\begin{equation}
\label{eq:phen}
R^{\textrm{(phen)}}(\tau,s_0,\beta,\hat m_q)=\frac{1}{\pi}\int_0^\infty e^{-s\tau}{\rm Im}\Pi^{\textrm{(phen)}}(s) \, ds
=R^{\textrm{(resonance)}}(\tau,\beta)+R^{\textrm{(ESC)}}(\tau,s_0,\hat m_q).
\end{equation}

Finally, the master equation for QCD sum rule can be obtained by demanding the equivalence between
Eq.~\eqref{eq:theoretical} and \eqref{eq:phen}:
\begin{equation}
\label{eq:master}
R^{\textrm{(theo)}}(\tau,\hat m_q)=R^{\textrm{(phen)}}(\tau,s_0,\beta,\hat m_q),
\end{equation}
which can be used to obtain the predictions for $s_0$, $\beta$ and $\hat m_q$ providing we take the
condensates and instanton size in the theoretical side as well as the physical
parameters for $\sigma$ and $f_0(980)$ in the phenomenological side as input parameters
\footnote{We can use Eq.~\eqref{eq:master} to obtain predictions for
resonance parameters in our phenomenological spectral density as in Ref.~\cite{Wang:2016sdt}
in principle. However, because the theoretical side of Eq.~\eqref{eq:master} is proportional to the
square of the light quark mass $m_q$, the master equation is sensitive with the value of $m_q$,
thus stable match between the two sides of the master equation is difficult to establish providing
different input $m_q$. Conversely, by taking the resonance parameters as input parameters, we can
use Eq.~\eqref{eq:master} to constrain the value of $m_q$ effectively.}.

Obviously, because of the truncation of OPE and the simplicity of the phenomenological
spectral density, Eq.~\eqref{eq:master} can not be valid for all $\tau$, thus one requires
a sum rule window in which the validity of the master equation can be established.
Benmerrouche et al.\ presented a method based on the H\"older inequality which provides
fundamental constraints on QCD sum rules \cite{Benmerrouche:1995qa}. By placing the
excited states and continuum contributions on the theoretical side, we obtain
\begin{equation}
\begin{split}
R^{\textrm{(theo-ESC)}}(\tau, s_0,\hat m_q)&\equiv
R^{\textrm{(theo)}}(\tau,\hat m_q)-R^{\textrm{(ESC)}}(\tau,s_0,\hat m_q)\\
&=\frac{1}{\pi}\int_{0}^{s_0} e^{-s\tau} \textrm{Im}\Pi^{\textrm{(phen)}}(s)\,ds,
\end{split}
\end{equation}
then the H\"older inequality for QCD sum rules can be written as
\begin{equation}
\label{eq:holder}
R^{\textrm{(theo-ESC)}}( \omega \tau_1+(1-\omega)\tau_2, s_0,\hat m_q)\leq
\left[R^{\textrm{(theo-ESC)}}(\tau_1,s_0,\hat m_q)\right]^\omega
\left[R^{\textrm{(theo-ESC)}}(\tau_2,s_0,\hat m_q)\right]^{1-\omega},
\end{equation}
where $0\leq \omega\leq 1$ and for parameters $\tau_1$ and $\tau_2$ we demand $\tau_1<\tau_2$.
Notice that different value of $\hat m_q$ does not change the allowed ($\tau$, $s_0$) region
from the H\"older inequality, thus we can set any value for $\hat m_q$ in Eq.~\eqref{eq:holder}.
Following Ref.~\cite{Benmerrouche:1995qa}, we will perform a local analysis on Eq.~\eqref{eq:holder}
with $\tau_2-\tau_1=\delta \tau=0.01\,\textrm{GeV}^{-2}$.

The only starting point of the H\"older inequality is that $\textrm{Im}\Pi^{\textrm{(phen)}}(s)$ should be
positive because of its relation to physical spectral functions, thus Eq.~\eqref{eq:holder}
must be satisfied if sum rules are to consistently describe integrated physical spectral functions.
In this paper, we will use the same iterative procedure to determine the sum rule window from the
H\"older inequality rigorously as in Ref.~\cite{Wang:2016sdt}, i.e., by choosing the maximally
allowed region $[\tau_{\textrm{min}}, \tau_{\textrm{max}}]$
of the H\"older inequality which is consistent with fitted $s_0$, where 
$\tau_{\textrm{min}}$ and $\tau_{\textrm{max}}$ are respectively the
lower bound and upper bound of the allowed $\tau$ region.

In order to match the two sides of the master equation \eqref{eq:master} in the sum rule window,
a weighted-least-squares method \cite{Leinweber:1995fn} will be used in this paper. By randomly
generating 200 sets of Gaussian distributed phenomenological input QCD parameters with given
uncertainties (10\% in this paper, which is the typical uncertainty in QCDSR) at
$\tau_j=\tau_\textrm{min}+(\tau_\textrm{max}-\tau_\textrm{min})\times(j-1)/(n_B-1)$, where
$n_B=21$, we can estimate the standard deviation $\sigma_{\textrm{theo}}(\tau_j)$ for
$R^{\textrm{(theo)}}(\tau_j,\hat m_q)$ \footnote{In practice, we will divide
$R^{(\textrm{theo})}$ by $\hat m^2_q$ in order to remove the to-be-fitted parameter
from the theoretical side, i.e., we estimate the standard deviation for
$R^{\textrm{(theo)}}(\tau_j,\hat m_q)/\hat m_q^2$.}. Then, the phenomenological output
parameters $s_0$, $\beta$ and $\hat m_q$ can be obtained by minimizing
\begin{equation}
\label{eq:chi2}
\chi^2=\sum_{j=1}^{n_B}\frac{(R^{\textrm{(theo)}}(\tau_j,\hat m_q)
-R^{\textrm{(phen)}}(\tau_j,s_0,\beta,\hat m_q))^2}{\sigma_{\textrm{theo}}^2(\tau_j)}.
\end{equation}

\section{Numerical results}%

In the numerical analysis, we use the central values of input QCD
parameters (at $\mu_0=1\,\textrm{GeV}$) as follows \cite{Narison:2014wqa,Schafer:1996wv}
\begin{equation}
\label{eq:input}
\begin{split}
&\langle \alpha_s G^2\rangle =0.07\,\textrm{GeV}^4,~
\langle m_q\bar qq\rangle =-(0.1\,\textrm{GeV})^4,\\
&\kappa \alpha_s\langle \bar qq\rangle^2=\kappa\times 1.49\times 10^{-4}\textrm{GeV}^6,~\rho=1/0.6\,\textrm{GeV}^{-1}.
\end{split}
\end{equation}
The size of $\kappa$ have been observed in different channels to be 2--4
\cite{Chung:1984gr,Narison:1995jr,Narison:2009vy}. Based on our previous study,
$\kappa=2.8$ is the favored result in the vector channel with a traditional ESC model \cite{Wang:2016sdt}.
Although the factorization violation effect may differ between channels, it is still reasonable
to assume the value of $\kappa$ is in the region of  2--3 in the scalar channel, too. 
Thus we consider $\kappa=2.0$ and $\kappa=3.0$ in our analysis, and as outlined below, we demonstrate
that $\kappa\sim 2$ leads to greater agreement between our light quark mass predictions and the PDG value.
In this paper, we will minimize the $\chi^2$ with 1000 sets of Gaussian distributed input QCD parameters listed in
Eq.~\eqref{eq:input} with 10\% uncertainties. Based on these 1000 fitting samples, we can obtain the
median and the asymmetric standard deviations from the median for all output parameters, thus we
obtain the uncertainty originating from uncertainties of QCD parameters for $s_0$, $\beta$ and
$\hat m_q$. \footnote{The mass and width of $\sigma$ and $f_0(980)$ will be considered as fixed
input parameters in each fit. However, we will input different combination of parameters for resonances
based on experiment to estimate the uncertainties for output parameters
which originate from parameters of resonances in the following.}

\begin{figure}[htbp]
\centering
\subfigure[~$\kappa=2.0$]{\label{fig:sr1}\includegraphics[scale=0.78]{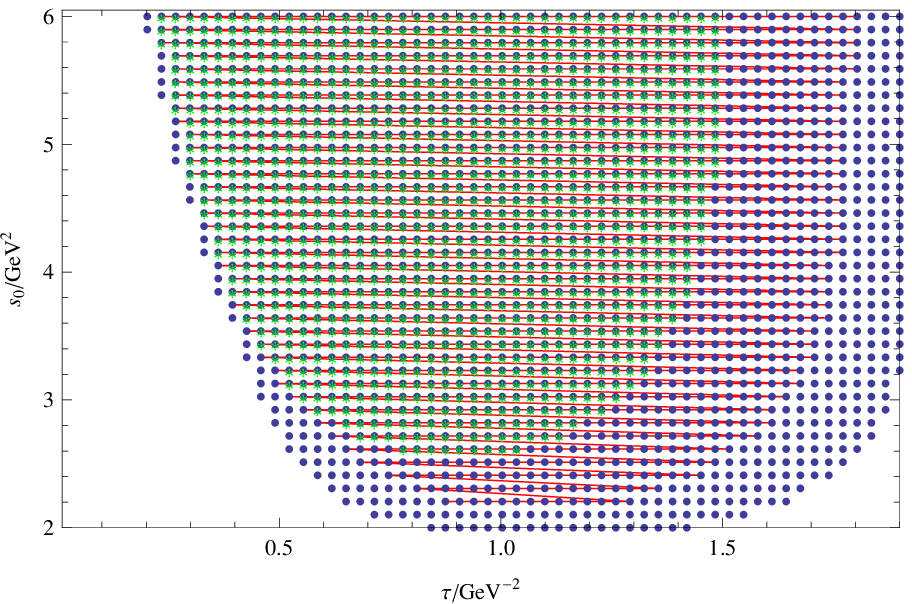}}
\subfigure[~$\kappa=3.0$]{\label{fig:sr2}\includegraphics[scale=0.78]{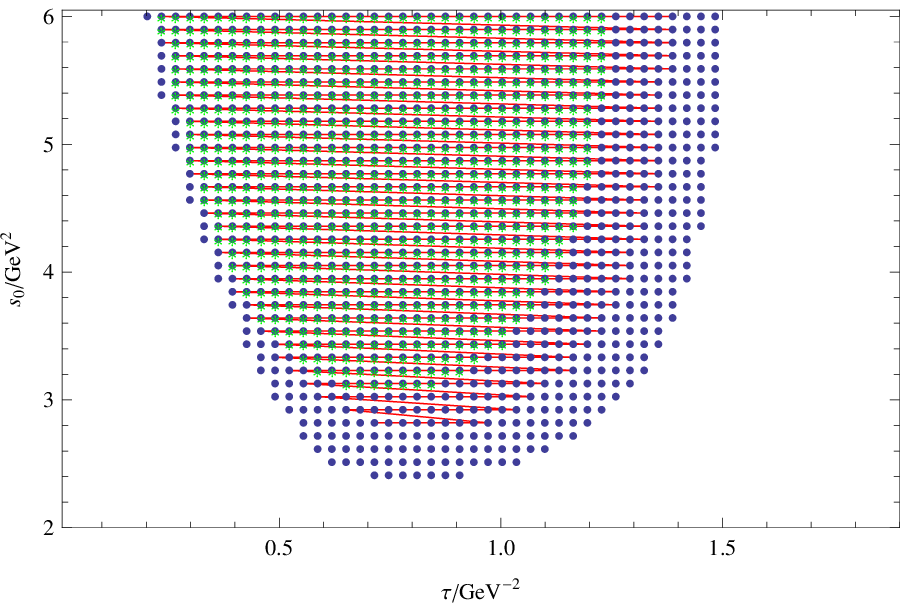}}
\caption{\label{fig:srwindow} The region allowed by the H\"older inequality for $\kappa=2.0$ (a) and $\kappa=3.0$ (b).
The region with (blue) dot or (red) line is allowed for
sum rule with or without instanton contribution respectively.
The region with (green) asterisk is allowed for sum rule without both $\alpha_s$
corrections to dimension-4 operators and instanton contribution. }
\end{figure}

In FIG.~\ref{fig:srwindow}, we plot the allowed region for ($\tau$, $s_0$) by the H\"older inequality for
$\kappa=2.0$ and $\kappa=3.0$ respectively. From this figure, we find that the $\alpha_s$ corrections to
$\langle\alpha_s G^2\rangle$ and $\langle m_q\bar qq\rangle$ extend the allowed region to a higher $\tau$ region and lower
$s_0$ region as in the $\rho$ channel \cite{Wang:2016sdt}, and the instanton contribution extends the allowed region further more.
Thus both $\alpha_s$ corrections to dimension-4 operators and instanton contribution are important 
since we adopt the same iterative procedure as described in Ref.~\cite{Wang:2016sdt} to determine
the sum rule window from the H\"older-inequality-allowed region rigorously.

Taking the experimental values of mass and width for $\sigma$ and $f_0(980)$ \cite{Olive:2016xmw}
\begin{equation}
\label{eq:resinput}
m_\sigma=400-550\,\textrm{MeV},~\Gamma_\sigma=400-700\,\textrm{MeV},~
m_{f_0}=990\pm20\,\textrm{MeV},~\Gamma_{f_0}=10-100\,\textrm{MeV}
\end{equation}
as our input in the phenomenological spectral density model, we obtain different fitted $s_0$, $\beta$
and $\hat m_q$ by minimizing the corresponding $\chi^2$ function. Detailed results are listed in
TABLE~\ref{tab:newfitresult} where we show the fitted results for $\kappa=2.0$
and $\kappa=3.0$ respectively. From this table we find that we can achieve very stable fits with $\kappa=2.0$,
all uncertainties of output parameters are less than 10\% providing 10\% uncertainties of input QCD parameters.
When we set $\kappa=3.0$, the uncertainty of $\hat m_q$ will reach to about 14\%-18\%, still in the accepted range of uncertainties for QCDSR.

\begin{table}[htbp]
\centering
\begin{tabular}{c|cccc|cccc}
  \hline
  \hline
&
\multicolumn{4}{c|}{Inputs}&\multicolumn{4}{c}{Outputs}\\
&
$m_\sigma$/MeV & $\Gamma_\sigma$/MeV & $m_{f_0}$/MeV & $\Gamma_{f_0}$/MeV & $s_0$/GeV$^2$ & $\beta$ & $\hat m_q$/MeV & $m_q(2\,\textrm{GeV})$/MeV\\
\hline
\multirow{8}{*}{$\kappa=2.0$}
&400&400&990&100& $2.77_{-0.16}^{+0.14}$ & $0.941_{-0.023}^{+0.016}$ & $7.02_{-0.44}^{+0.62}$ &$4.0_{-0.3}^{+0.4}$\\%
&400&400&990& 10& $2.71_{-0.15}^{+0.13}$ & $0.995_{-0.002}^{+0.001}$ & $6.87_{-0.40}^{+0.54}$ &$4.0_{-0.2}^{+0.3}$\\%
&400&700&990&100& $2.77_{-0.16}^{+0.14}$ & $0.955_{-0.020}^{+0.013}$ & $6.40_{-0.41}^{+0.58}$ &$3.7_{-0.2}^{+0.3}$\\%
&400&700&990& 10& $2.71_{-0.15}^{+0.13}$ & $0.996_{-0.002}^{+0.001}$ & $6.25_{-0.37}^{+0.50}$ &$3.6_{-0.2}^{+0.3}$\\%
&550&400&990&100& $2.66_{-0.20}^{+0.16}$ & $0.935_{-0.033}^{+0.024}$ & $8.41_{-0.51}^{+0.70}$ &$4.8_{-0.3}^{+0.4}$\\%
&550&400&990& 10& $2.60_{-0.19}^{+0.15}$ & $0.995_{-0.003}^{+0.002}$ & $8.35_{-0.49}^{+0.67}$ &$4.8_{-0.3}^{+0.4}$\\%
&550&700&990&100& $2.73_{-0.16}^{+0.14}$ & $0.958_{-0.024}^{+0.016}$ & $7.16_{-0.44}^{+0.62}$ &$4.1_{-0.3}^{+0.4}$\\%
&550&700&990& 10& $2.68_{-0.16}^{+0.14}$ & $0.996_{-0.002}^{+0.001}$ & $7.06_{-0.42}^{+0.57}$ &$4.1_{-0.2}^{+0.3}$\\%
\hline
\multirow{8}{*}{$\kappa=3.0$}
&400&400&990&100& $3.03_{-0.09}^{+0.09}$ & $0.872_{-0.078}^{+0.030}$ & $9.00_{-0.80}^{+1.60}$ &$5.2_{-0.5}^{+0.9}$\\%
&400&400&990& 10& $2.94_{-0.09}^{+0.08}$ & $0.990_{-0.005}^{+0.002}$ & $8.56_{-0.69}^{+1.25}$ &$4.9_{-0.4}^{+0.7}$\\%
&400&700&990&100& $3.03_{-0.09}^{+0.09}$ & $0.896_{-0.071}^{+0.026}$ & $8.28_{-0.75}^{+1.54}$ &$4.8_{-0.4}^{+0.9}$\\%
&400&700&990& 10& $2.95_{-0.08}^{+0.08}$ & $0.992_{-0.004}^{+0.002}$ & $7.82_{-0.63}^{+1.14}$ &$4.5_{-0.4}^{+0.7}$\\%
&550&400&990&100& $2.99_{-0.10}^{+0.10}$ & $0.835_{-0.095}^{+0.041}$ & $10.7_{-0.9}^{+1.6}$   &$6.2_{-0.5}^{+0.9}$\\%
&550&400&990& 10& $2.90_{-0.10}^{+0.09}$ & $0.986_{-0.008}^{+0.003}$ & $10.5_{-0.8}^{+1.5}$   &$6.0_{-0.5}^{+0.9}$\\%
&550&700&990&100& $3.03_{-0.09}^{+0.09}$ & $0.881_{-0.082}^{+0.032}$ & $9.27_{-0.81}^{+1.61}$ &$5.3_{-0.5}^{+0.9}$\\%
&550&700&990& 10& $2.95_{-0.08}^{+0.08}$ & $0.991_{-0.005}^{+0.002}$ & $8.91_{-0.71}^{+1.29}$ &$5.1_{-0.4}^{+0.7}$\\%
  \hline
  \hline
\end{tabular}
\caption{\label{tab:newfitresult} Fitted results with different choices of the mass and width for the two resonances.
All uncertainties of QCD input parameters listed in Eq.~\eqref{eq:input} are set to 10\%.}
\end{table}

The suggested light quark mass at 2\,GeV from PDG reads \cite{Olive:2016xmw}
\begin{equation}
\label{eq:pdgvalue}
m_q^{\textrm{PDG}} (2\,\textrm{GeV})=\frac{1}{2}(m_u+m_d)=3.5^{+0.7}_{-0.3}\,\textrm{MeV}.
\end{equation}
To compare our fitted results with $m_q^{\textrm{PDG}} (2\,\textrm{GeV})$, we also list the corresponding
light quark mass at 2\,GeV from our fitting procedure in TABLE~\ref{tab:newfitresult}.
Based on these data, we can obtain
\begin{equation}
m_q(2\,\textrm{GeV})=4.1\pm0.4\textrm{(resonance)}^{+0.4}_{-0.3}\textrm{(QCD)}\,\textrm{MeV}=4.1^{+0.6}_{-0.5}\,\textrm{MeV}
\end{equation}
for $\kappa=2.0$ and
\begin{equation}
m_q(2\,\textrm{GeV})=5.3\pm0.6\textrm{(resonance)}^{+0.8}_{-0.5}\textrm{(QCD)}\,\textrm{MeV}=5.3^{+1.0}_{-0.8}\,\textrm{MeV}
\end{equation}
for $\kappa=3.0$, where we report the average value
of $m_q(2\,\textrm{GeV})$ with different resonance parameters, and combine the standard deviation
and the asymmetric standard deviation which originate from different resonance parameters and uncertainties
of QCD input parameters respectively. 
Comparison with the PDG tends to favor the smaller value of $\kappa$. However, 
since an exact value of $\kappa$ not known, we use the average
value for $\kappa=2.0$ and $\kappa=3.0$ as a conservative determination of our final result
\begin{equation}
m_q(2\,\textrm{GeV})=4.7^{+0.8}_{-0.7}\,\textrm{MeV}.
\end{equation}
This central value result is slightly heavier than the PDG value in Eq.~\eqref{eq:pdgvalue}, however,
is still consistent with it. 
We expect further experimental data on the mass and width for
$\sigma$ and $f_0(980)$ would reduce the uncertainty for our prediction.

From TABLE~\ref{tab:newfitresult}, we also can obtain
\begin{equation}
s_0=2.70\pm0.06\textrm{(resonance)}^{+0.14}_{-0.17}\textrm{(QCD)}\,\textrm{GeV}^2=2.70^{+0.15}_{-0.18}\,\textrm{GeV}^2
\end{equation}
for $\kappa=2.0$ and
\begin{equation}
s_0=2.98\pm0.05\textrm{(resonance)}^{+0.09}_{-0.09}\textrm{(QCD)}\,\textrm{GeV}^2=2.98^{+0.10}_{-0.10}\,\textrm{GeV}^2
\end{equation}
for $\kappa=3.0$.

We notice that the uncertainty of the fitted continuum threshold $s_0$ is astonishing small, especially those originating from different
resonance parameters. Krasnikov et al. pointed out that contributions from below the $n$-th resonances and from above the $n+1$-th resonances
in the spectral density can be separated by using $s_0=\frac{1}{2}(m_n^2+m_{n+1}^2)$ where $m_n$ and $m_{n+1}$ is the mass of the $n$-th and
$n+1$-th resonance respectively \cite{Krasnikov:1981vw}, i.e., $s_0$ is determined only by the mass positions
of the two nearest resonances in the spectral density which are located at the two sides of $s_0$. 
If this choice for $s_0$ is also applicable in the present case, then we can
 give a simple explanation why $s_0$ is not affected a lot by different resonance parameters:
although we input different mass and width for $\sigma$ and different width for $f_0(980)$,
the mass of $f_0(980)$ is fixed, thus
\begin{equation}
\label{eq:continuum}
s_0=\frac{1}{2}(m_2^2+m_3^2)
\end{equation}
will not change significantly during our fitting procedure, where $m_3$ is the next excited state in the present
scalar channel which couples with the scalar current $j_s$ strongly.
By using $m_2=990\,\textrm{MeV}$ from experiment and Eq.~\eqref{eq:continuum}, we can estimate the mass for the next resonance,
which ranges from 2.10\,GeV ($\kappa=2.0$) to 2.23\,GeV ($\kappa=3.0$). Based on the average value of $m_3$ which is about 2.17\,GeV,
 $f_0(1370)$, $f_0(1500)$ and $f_0(1710)$ are sufficiently weakly-coupled to $j_s$ to be negligible.
On the other hand, our result favors one resonance in the group of $f_0(2020)$, $f_0(2100)$, $f_0(2200)$ and $f_0(2330)$ (which are all $0^+(0^{++})$ resonances
listed in the latest Review of Particle Physics \cite{Olive:2016xmw}) for an appreciable coupling to 
$j_s$ and the exponential suppression in the Laplace sum-rule enables inclusion within the continuum.  

The continuum threshold $s_0$ is introduced to separate out the contributions from excited states and continuum in the phenomenological spectral density.
This expected purpose is achieved in many works of QCD sum rules under the narrow resonance approximation. However,
we deal with resonances with non-zero width in the present case. Thus there is a second possibility that we actually cannot
separate the ESC contributions from the first several resonances contributions exactly because of the overlapping contributions from different resonances.
If this is the case, then the traditional one parameter (i.e., $s_0$) ESC model is too simple to describe the true physical spectral density.
Although a large $s_0$ is obtained during the fitting procedure, which leads to $\sqrt{s_0}=1.64-1.73\,\textrm{GeV}$, 
we still cannot conclude that those scalar mesons between 1-2\,GeV are excluded from
the phenomenological spectral density. 
But luckily, due to the heavier mass and relative small two-pion decay branching ratio, the contributions 
from $f_0(1370)$ and $f_0(1500)$ are expected to be very small. 
For $f_0(1370)$ as an example, if we assume that there is a contribution from $f_0(1370)$ to the scalar form factor $F_s$, which
has the same magnitude of contribution as $f_0(980)$ (obviously, the magnitude of $f_0(1370)$ is overestimated
because the position of $f_0(1370)$ is further away from the normalization point of $F_s$, i.e., $s=0$, 
than $f_0(980)$), then we can estimate a rough relative contribution from
$f_0(1370)$ and $f_0(980)$ to the Borel-transformed correlation function in the whole sum rule window, 
which is about 20-30\%. After considering the relative small two-pion decay branching ratio, the contribution from $f_0(1370)$ 
to the Borel-transformed correlation function will be at most at the same magnitude of the uncertainty of QCDSR.
Thus the fitted light quark mass will not be affected a lot after including these contributions. However,
to solve the $s_0$ problem comprehensively and rigorously, a better description of the ESC is deserved, which needs further studies.

By extracting the coefficients for the two standard Breit-Wigner functions in the
phenomenological spectral density in Eq.~\eqref{eq:model1}, we can define
two effective coupling constants which describe the coupling between the scalar
current $j_s$ and the two resonances ($\sigma$ and $f_0(980)$) as follows
\begin{gather}
\label{eq:F1}
\lambda_\sigma=\beta \frac{3}{64\pi}m_\pi^4 (m_\sigma^2+\Gamma_\sigma^2)\frac{m_\sigma}{\Gamma_\sigma},\\
\label{eq:F2}
\lambda_{f_0}=(1-\beta) \frac{3}{64\pi}m_\pi^4 (m_{f_0}^2+\Gamma_{f_0}^2)\frac{m_{f_0}}{\Gamma_{f_0}}.
\end{gather}

These two effective coupling constants can be related to other physical quantities. By inserting
one-particle intermediate states ($\sigma$ and $f_0(980)$ states) as part of a complete set, $\int\frac{d^4\bm{k}}{(2\pi)^3 2 E_{\bm{k}}}
(|\sigma(k)\rangle\langle\sigma(k)|+|f_0(980)(k)\rangle\langle f_0(980)(k)|)
+$ ``other intermediate states'', into the correlation function \eqref{eq:correlator},
 a traditional phenomenological
density can be obtained \footnote{We have extended narrow resonances
model with Breit-Wigner resonances model for $\sigma$ and $f_0(980)$.}
\begin{equation}
\label{eq:model2}
\frac{1}{\pi}{\rm Im}\Pi^{\textrm{(phen)}}(s)=m_q^2 f_\sigma^2 m_\sigma^2 \cdot
\frac{1}{\pi}\frac{m_\sigma \Gamma_\sigma}{(s-m_\sigma^2)^2+m_\sigma^2\Gamma_\sigma^2}
+ m_q^2 f_{f_0}^2m_{f_0}^2 \cdot\frac{1}{\pi}\frac{m_{f_0} \Gamma_{f_0}}{(s-m_{f_0}^2)^2+m_{f_0}^2\Gamma_{f_0}^2}
+\frac{1}{\pi}{\rm Im}\Pi^{\textrm{(ESC)}}(s),
\end{equation}
where $f_\sigma$ and $f_{f_0}$ are the decay constants of $\sigma$ and $f_0(980)$ respectively,
which satisfy $\langle 0|\frac{1}{\sqrt{2}}(\bar uu+\bar dd)|\sigma\rangle=f_\sigma m_\sigma$
and $\langle 0|\frac{1}{\sqrt{2}}(\bar uu+\bar dd)|f_0(980)\rangle=f_{f_0} m_{f_0}$.
Comparing Eq.~\eqref{eq:model1} with \eqref{eq:model2}, we can connect our effective
coupling constants with $f_\sigma$ and $f_{f_0}$ as follows
\begin{gather}
\lambda_\sigma=m_q^2(\mu) f_\sigma^2(\mu) m_\sigma^2,\\
\lambda_{f_0}=m_q^2(\mu) f_{f_0}^2(\mu) m_{f_0}^2,
\end{gather}
where $\mu$ is an energy scale.

\begin{table}[htbp]
\centering
\begin{tabular}{c|cccc|cccc}
  \hline
  \hline
&$m_\sigma$/MeV & $\Gamma_\sigma$/MeV & $m_{f_0}$/MeV & $\Gamma_{f_0}$/MeV
& $\lambda_\sigma$/10$^{-6}$GeV$^6$ & $f_\sigma(2\,\textrm{GeV})$/GeV
& $\lambda_{f_0}$/10$^{-6}$GeV$^6$ & $f_{f_0}(2\,\textrm{GeV})$/GeV \\
\hline
\multirow{8}{*}{$\kappa=2.0$}
&400&400&990&100& 1.68&0.81&3.22&0.45 \\%
&400&400&990& 10& 1.77&0.83&2.70&0.42 \\%
&400&700&990&100& 1.98&0.95&2.46&0.43 \\%
&400&700&990& 10& 2.06&1.00&2.16&0.41 \\%
&550&400&990&100& 3.31&0.69&3.55&0.40 \\%
&550&400&990& 10& 3.52&0.71&2.70&0.35 \\%
&550&700&990&100& 3.32&0.81&2.29&0.37 \\%
&550&700&990& 10& 3.45&0.82&2.16&0.36 \\%
\hline
\multirow{8}{*}{$\kappa=3.0$}
&400&400&990&100& 1.55&0.60&6.99&0.51 \\%
&400&400&990& 10& 1.76&0.68&5.41&0.48 \\%
&400&700&990&100& 1.85&0.71&5.68&0.50 \\%
&400&700&990& 10& 2.05&0.80&4.32&0.47 \\%
&550&400&990&100& 2.96&0.50&9.01&0.49 \\%
&550&400&990& 10& 3.49&0.57&7.57&0.46 \\%
&550&700&990&100& 3.06&0.60&6.50&0.49 \\%
&550&700&990& 10& 3.44&0.66&4.86&0.44 \\%
  \hline
  \hline
\end{tabular}
\caption{\label{tab:decayconstant} Effective coupling constants and decay constants of $\sigma$ and $f_0(980)$.}
\end{table}

In TABLE~\ref{tab:decayconstant}, we list the effective coupling constants and the decay constants
of $\sigma$ and $f_0(980)$ based on our fitted results listed in TABLE~\ref{tab:newfitresult}.
For simplicity, we only use the central values of the fitted $\beta$ and $m_q(2\,\textrm{GeV})$
to estimate the effective coupling constants and the decay constants, and we do not estimate
the uncertainties for these constants. Based on our estimation, we obtain the average value
$\bar f_\sigma(2\,\textrm{GeV})=0.83$\,GeV for $\kappa=2.0$ and
$\bar f_\sigma(2\,\textrm{GeV})=0.64$\,GeV for $\kappa=3.0$, we may conclude that
the value of the decay constant of $\sigma$ at 2\,GeV is around $0.64-0.83\,\textrm{GeV}$.
In Ref.~\cite{Celenza:1992vz}, Celenza et al. estimated the value of $f_\sigma$ by using
the Nambu-Jona-Lasinio (NJL) model, their result reads $f_\sigma(2\,\textrm{GeV})=0.42\,\textrm{GeV}$,
0.48\,GeV, 0.35\,GeV and 0.43\,GeV depending on different model parameters \footnote{We have
converted the value of $f_\sigma$ at the momentum cutoff in the NJL model into the value of $f_\sigma$ at 2\,GeV.}.
Our result which favors a larger coupling between $j_s$ and the $\sigma$ state is more consistent with
the result from the linear sigma model ($L\sigma M$), which gives
$f_\sigma(2\,\textrm{GeV})=0.65-0.90\,\textrm{GeV}$ \cite{Clement:1991sh}
\footnote{We use the result $\langle 0|m_q(\bar uu+\bar dd)|\sigma\rangle
=f_\pi m_\pi^2$ from the linear sigma model, where $f_\pi=93\,\textrm{MeV}$ is the pion decay constant,
$m_q^{\textrm{PDG}}(2\,\textrm{GeV})$ and the mass
of $\sigma$ from experiment to estimate $f_\sigma(2\,\textrm{GeV})$.}.
We also obtain $\bar f_{f_0}(2\,\textrm{GeV})=0.40$\,GeV for $\kappa=2.0$ and
$\bar f_{f_0}(2\,\textrm{GeV})=0.48$\,GeV for $\kappa=3.0$, thus the value of the decay constant
of $f_0(980)$ at 2\,GeV is about $0.40-0.48\,\textrm{GeV}$.
It is interesting that our $f_0(980)$ decay constant agrees with Ref.~~\cite{Cheng:2005ye}, where $f_{f_0}(1\,\textrm{GeV})\simeq0.35\,\textrm{GeV}$ and
$f_{f_0}(2.1\,\textrm{GeV})\simeq0.41\,\textrm{GeV}$ considering the differences in our approaches. 

We also tried to use a one resonance model, i.e., set $\beta=0$ or 1 in Eq.~\eqref{eq:peak},
to finish our fitting procedure. However, after including the constraint on the
phenomenological spectral density from low-energy theorem, i.e., $|F_s(0)|^4=m_\pi^4$,
none of the combination of resonance mass and width based on Eq.~\eqref{eq:resinput}
would lead to reasonable match between the two sides of the QCDSR master equation \eqref{eq:master}
in the QCD sum rule window allowed by the H\"older inequality. A simple explanation of
this astonishing result is that the scalar form factor does receive contributions both
from $\sigma$ and $f_0(980)$ as we conjectured in the previous section. 

Based on the above results which lead to $\beta\sim 1$, it seems that the $\sigma$ peak dominate the 
resonance contributions in the phenomenological spectral density, however, this expectation is not necessarily true because of the
large gap between the peaks of $\sigma$ and $f_0(980)$. Although contribution from $\sigma$ peak dominates 
the low $s$ region in the phenomenological spectral density, there is also a significant 
contribution from the $f_0(980)$ peak in the whole sum rule window. 
In fact, the total contribution from the $\sigma$ peak to the
Borel-transformed correlation function in the sum rule window,
i.e., $\int_{\tau_{\textrm{min}}}^{\tau_{\textrm{max}}} R^{(\sigma\textrm{peak})}(\tau) d\tau$, can be about 46\%-
65\% of total contributions from both the $\sigma$ and $f_0(980)$ peaks with $\kappa=2.0$. The specific percent changes as we
input different mass and width parameters for the two resonances. For larger vacuum factorization violation factor, 
the contribution from $\sigma$ will reduce. However, the existence
of the enigmatic $\sigma$ is still essential in our procedure with $\kappa=3.0$.

Finally, the effects of $\alpha_s$ corrections to dimension-4 operators and instanton
contribution are also studied. From FIG.~\ref{fig:srwindow} we have learned that without
these effects, the allowed $\tau$-$s_0$ region would shrink, thus it is more difficult to
obtain acceptable fitted result which is consistent with the H\"older inequality. In fact,
we cannot obtain stable fit with $\kappa=3.0$ without these effects, and with $\kappa=2.0$,
we would obtain fitted-$\hat m_q$ (and $m_q(2\,\textrm{GeV})$) which is significantly larger than
the physical value from PDG. Based on these results, we can conclude that both the $\alpha_s$
corrections to dimension-4 operators and the instanton contribution are essential contribution
in the theoretical representation of the correlation function \eqref{eq:correlator}.

\section{Conclusions}

In this paper, we have constructed a phenomenological spectral density model with two
Breit-Wigner type resonances ($\sigma$ and $f_0(980)$) for the $I=0$ scalar channel with a normalization constrained by the ChPT low-energy theorem, 
and conducted the sum rule analysis of this channel in the H\"older-inequality-determined
sum rule window via the Monte-Carlo based fitting procedure. Based on our analysis, we
obtain a prediction for the light quark mass $m_q$ using the experimental results for the
masses and widths of $\sigma$ and $f_0(980)$. The agreement between our result
$m_q(2\,\textrm{GeV})=4.7^{+0.8}_{-0.7}\,\textrm{MeV}$, the PDG value, and QCDSR determinations 
in the pion channel provide a consistent framework connecting QCD and low-energy hadronic physics (see also Ref.~\cite{Fariborz:2015vsa}).   
Furthermore, this agreement in the quark mass determinations confirms
the validity of our improved Monte-Carlo based QCD sum rules, which has previously been
systematically examined in the $\rho$ meson channel in Ref.~\cite{Wang:2016sdt}. 
Our results indicate both $\sigma$ and $f_0(980)$ couple to the scalar current $j_s$ strongly, i.e.,
both $\sigma$ and $f_0(980)$ have $\bar qq$-component. 

The continuum threshold $s_0$ obtained from our fitting procedure, seems to exclude scalar 
mesons between 1-2\,GeV from the ESC contributions. There are two possibilities to understand this result. 
One possibility is that those mesons are weakly-coupled enough to be excluded from the phenomenological spectral density, and we expect the 
next excited state is in the group of scalar mesons which is heavier than 2\,GeV and the exponential 
suppression in the Laplace sum-rule enables inclusion within the continuum. 
The other possibility is that the traditional ESC model is too simple to describe the 
true ESC contributions exactly, and we cannot use one parameter to separate ESC contributions from a spectral density 
with overlapping resonance contributions, thus a more realistic ESC model includes parameters other than $s_0$ is needed 
to solve this problem comprehensively and rigorously. 

From our analysis, we also obtain the
value of the decay constants of $\sigma$ and $f_0(980)$ at 2\,GeV, which are respectively around
$0.64-0.83$\,GeV and around $0.40-0.48$\,GeV. These two decay constants can be used in further studies on
the decays of heavier mesons, e.g., B mesons, which can decay through the $s$-wave two pions state.

\begin{acknowledgments}
This work is supported by NSFC under grant 11175153, 11205093 and 11347020, and supported by Open Foundation of the Most Important
Subjects of Zhejiang Province, and K. C. Wong Magna Fund in Ningbo University.
TGS is supported by the Natural Sciences and Engineering Research Council of Canada (NSERC). Z. F. Zhang thanks
the University of Saskatchewan for its warm hospitality.
\end{acknowledgments}


\end{document}